# Construction of Attenuation Relationship of Peak Ground Velocity Using Machine Learning and Examination of Station Correction Factor

Wu Junjie[1], Yoshihisa Maruyama [2] and Wen Liu[2]

[1]Dept. of Urban Environment Systems,
Graduate School of Engineering, Chiba University
E-mail: ayka5587@chiba-u.jp
[2]Dept. of Urban Environment Systems,
Graduate School of Engineering, Chiba University

This study tries to develop new attenuation relationships of peak ground velocity using machine learning methods; random forest and neural network. In order to compare with the predictors obtained by machine learning, we have also constructed a new attenuation relationship of peak ground velocity using three-stage regression procedure proposed by Molas and Yamazaki (1995). In this study, 6,944 ground motion records at 1,184 seismic observation stations which were observed during the 32 earthquakes are employed to construct the attenuation relationships. Ground motion records from the 4 recent earthquakes are used as the test dataset. The test results show that when the shortest distance from the fault is small, the predictions by machine learning techniques are more accurate than the traditional equation. However, there is still a problem of overestimation in the predictors of machine learning, even if weights are added to the training dataset. In addition, the station correction factors based on machine learning were derived and proved to be correlated with the average shear wave velocity.

*Key Words :* *attenuation relationship, machine learning, random forest, neural network, station correction factors*

## 1. INTRODUCTION

The attenuation relationship is a method to predict the ground motion intensity of earthquake that may occur in the future based on the ground motion records of past earthquakes. The attenuation relationships are used in both deterministic and probabilistic seismic hazard analyses. The attenuation refers to the phenomenon that the farther away from the epicenter, the weaker the earthquake intensity. The previous attenuation relationships are empirical equations that predict the level of ground shaking, based on the source characteristics (e.g., earthquake magnitude), the propagation path (e.g., the shortest distance from the fault), and the local site conditions, etc. In the United States, starting with the pioneering research of Gutenberg and Richter (1942), research on attenuation relationship has been active until now, mainly for the purpose of seismic risk evaluation and strong motion prediction near faults.

In Japan, apart from the research on the seismic intensity and distance of Kawasumi (1943) and the research on the magnitude of Tsuboi (1954), the semi-empirical equation of the seismic motion characteristics of Kanai (1957) is the first research on the attenuation relationship. As the development of statistical analysis methods and more ground motion records are obtained, the research of attenuation relationship has been greatly developed. However, due to the lack of ground motion records near the epicenter, it was found that previous attenuation relationships have low reliability at close range. Therefore, this study tries to develop new attenuation relationships of peak ground velocity (PGV) using machine learning methods; random forest and neural network.

Machine learning has become a large field of study that overlaps with many areas. The focal point of machine learning is learning, that is, acquiring knowledge from data. Specifically, it learns from and makes predictions based on data. Therefore, the large amount of ground motion data obtained offer the opportunity to develop new attenuation relationships using machine learning. In this study, random forest and neural network are used to predict the PGV. Previous studies have constructed attenuation relationships using random forest (Kubo 2018) and neural network (Derras 2012). In this study, we want to compare the predictors obtained by machine learning with a traditional one. Therefore, we have constructed a new attenuation equation of PGV using three-stage regression procedure proposed by Molas and Yamazaki (1995).

Tabel 1. The list of earthquake events used as the training data.

| No. | Date | Mw | Depth | Number of Records |
|---|---|---|---|---|
| 1 | 1997.03.26 | 6.0 | 8 | 58 |
| 2 | 1997.05.13 | 5.9 | 8 | 52 |
| 3 | 2000.10.16 | 6.6 | 11 | 245 |
| 4 | 2000.10.31 | 5.4 | 44 | 141 |
| 5 | 2001.03.24 | 6.9 | 51 | 274 |
| 6 | 2001.04.25 | 5.4 | 42 | 93 |
| 7 | 2003.05.26 | 7.0 | 71 | 392 |
| 8 | 2003.07.26 | 6.2 | 12 | 194 |
| 9 | 2003.09.26 | 8.0 | 42 | 342 |
| 10 | 2004.09.05 | 7.4 | 44 | 369 |
| 11 | 2004.10.23 | 6.5 | 13 | 294 |
| 12 | 2004.10.27 | 5.8 | 12 | 214 |
| 13 | 2004.11.29 | 6.8 | 48 | 180 |
| 14 | 2004.12.14 | 5.9 | 9 | 70 |
| 15 | 2005.03.20 | 6.6 | 9 | 192 |
| 16 | 2005.07.23 | 5.8 | 73 | 193 |
| 17 | 2005.08.16 | 7.1 | 42 | 395 |
| 18 | 2006.04.21 | 5.6 | 7 | 66 |
| 19 | 2006.05.02 | 5.1 | 15 | 64 |
| 20 | 2006.06.12 | 5.9 | 146 | 221 |
| 21 | 2006.08.31 | 4.8 | 76 | 107 |
| 22 | 2007.03.25 | 6.7 | 11 | 234 |
| 23 | 2007.07.16 | 6.7 | 17 | 304 |
| 24 | 2008.05.08 | 6.9 | 51 | 202 |
| 25 | 2008.06.14 | 6.9 | 8 | 250 |
| 26 | 2008.09.11 | 6.8 | 31 | 144 |
| 27 | 2009.08.11 | 6.2 | 23 | 287 |
| 28 | 2010.02.27 | 6.7 | 37 | 8 |
| 29 | 2011.03.09 | 7.3 | 8 | 234 |
| 30 | 2011.03.11 | 9.0 | 24 | 685 |
| 31 | 2011.04.11 | 6.6 | 6 | 306 |
| 32 | 2011.04.12 | 6.4 | 26 | 134 |

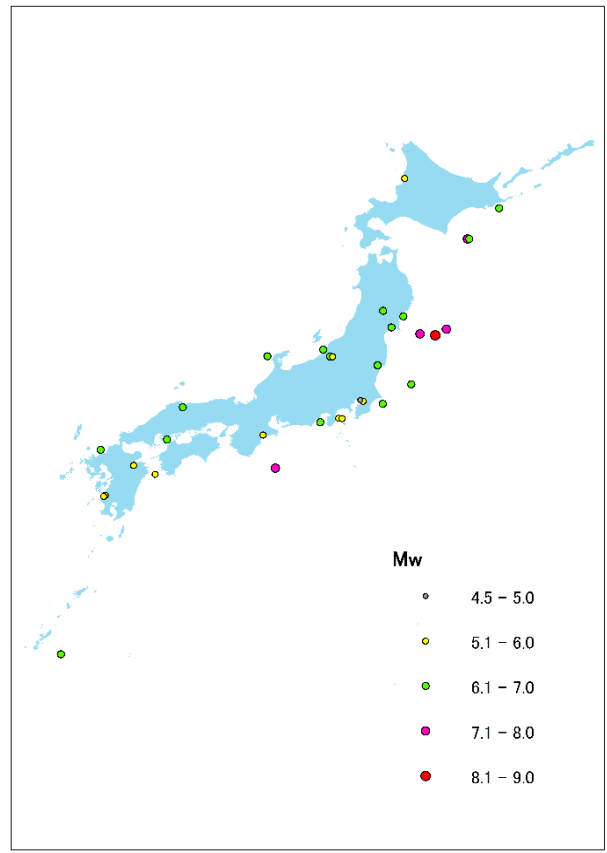

Fig. 1. Locations of epicenter and the moment magnitude of the 32 earthquake events.

## 2. DATA And METHODS

### 2.1 DATA

In this study, we use ground motion data obtained by K-NET (Kyoshin network) and KiK-net (Kiban Kyoshin network), which are strong-motion seismograph networks constructed by the National Research Institute for Earth Science and Disaster prevention (NIED). K-NET and KiK-net consist of more than 1,700 observation stations that reliably record strong ground motions. In this study, 6,944 ground motion records at 1,184 K-NET and KiK-net seismic observation stations which were observed during the 32 earthquakes are employed as training data to construct the attenuation relationships. Table 1 shows the list of earthquake events used as training data. The training dataset consists of earthquake events from 1997 to 2011. Figure 1 shows the epicenter locations and the moment magnitude. In order to focus on the prediction of close-range data, we give the following weights to the training data when constructing the attenuation equation, 0 - 25 km is 8, 25 km - 50 km is 4, 50 km - 100 km is 2, larger than 100 km is 1. Considering that all the 3 methods use the same explanatory variables and objective variable, we also give the same weights to the training data of the machine learning models. Moreover, ground motion data observed during the 4 recent earthquake events are used as the test data to evaluate the performances of predictors, including the April 14 foreshock of 2016 Kumamoto earthquake, the April 16 mainshock of 2016 Kumamoto earthquake, the 2018 Osaka earthquake and the 2018 Hokkaido Eastern Iburi earthquake.

Most of the attenuation relationships are empirical equations developed from a set of ground motion data. They estimate ground motion indices like PGV and the peak ground acceleration (PGA) as functions of source characteristics (e.g., earthquake magnitude), the propagation path (e.g., the shortest distance from the fault), and the local site conditions. Therefore, PGV is used as the objective variable, and the moment magnitude ($M_w$), the shortest distance from the fault ($r$), the earthquake source depth ($H$), and the dummy variable ($S_i$), which is mentioned later, are used as the explanatory variables to construct the attenuation relationships.

## 2.2 METHODS
(1) Regression analysis

In order to compare with the predictors obtained by machine learning, we have constructed a new attenuation equation of PGV using three-stage regression procedure proposed by Molas and Yamazaki (1995).

The first step determines the coefficients of the regression model given by Eq. 1 and the coefficients calculated serve as initial estimates.

$$log\ PGV = b_0 + b_1 M_w + b_2 r + b_3 log(r + c_1 * 10^{c_2 * M_w}) + b_4 H + C_i \tag{1}$$

Where PGV is the peak ground velocity, $M_w$ is the moment magnitude, $r$ is the shortest distance from the fault, $H$ is the earthquake source depth, $C_i$ is the station correction factor for station $i$ and $b_i$s are the coefficients to be determined.

The second step is the multilinear regression shown in Eq. 2.

$$log\ PGV = \sum_{j=1}^{K} a_j A_j + b_2 r + b_3 log(r + c_1 * 10^{c_2 * M_w}) + b_4 H + C_i \tag{2}$$

Where $A_j = 1$ for $j$th earthquake event (0 otherwise). In this step, $b_4 H$ and $C_i$ are constrained to the values determined in the first step. The distance dependence of the attenuation is then determined.

The third step is the regression shown in Eq. 3.

$$a_j = b_0 + b_1 M_w \tag{3}$$

Where $a_j$ is determined in the second step. This step determines the magnitude dependence of the attenuation relationship[9].

The first step is then repeated, except that $b_1$ to $b_3$ are constrained to the values from the second step and the third step. The cycle is repeated until the coefficients stabilize.

(2) Random forest

Random forest is an ensemble model consisting of many decision trees. Predictions are made by averaging the predictions from each decision tree. Alternatively, as a forest is a collection of trees, a random forest model is a collection of decision trees. The core idea behind random forest is to generate multiple decision trees from random subsets of the dataset. The following shows the algorithm of random forest.

I. Pick *N* records randomly from the dataset.
II. Build a decision tree consisting *M* features based on these *N* records.
III. Decide the number of trees and repeat I and II.
IV. For a new record, each decision tree predicts a value of output. The final output of random forest can be calculated by taking the average of all the values predicted by all the decision trees.

In this random forest model, the objective variable is PGV, and the moment magnitude ($M_w$), the shortest distance from the fault ($r$), the earthquake source depth ($H$) and the 1,184 dummy variables ($S_i$) are used as the explanatory variables. The dummy variables, $S_i$, are configured such that the mean of station correction factor is zero. For the *j*th data recorded at station *k*, $S_{i=k,j} = 1$ and $S_{i \neq k,j} = 0$ except if the data is recorded at the last (1184th) station, then $S_i$ is taken as -1 for *i*=1 to 1,183.

In this study, we use the Scikit-learn, which is a free software machine learning library for the Python programming language, to construct the random forest model. The main parameters to adjust when using Scikit-learn RandomForestRegressor are *n_estimators*, *max_depth* and *max_features*. The first one is the number of trees in the forest. The larger is the better. The second one is the maximum depth of the tree. The deeper tree has the more splits, and it captures more information about the data. The last one is the size of the random subsets of features to consider when splitting a node. The lower produces the less variance, but also the greater increases bias. These 3 parameters are determined by cross-validation.

(3) Neural network

A neural network model, more properly referred to as artificial neural network (ANN), is a forecasting method based on simple mathematical model of the brain[4]. It allows complex nonlinear relationships between the response variable and its predictions. The design of the neural network model requires several choices concerning the selection of the input nodes that are relevant to the output, the size of hidden layers, and the functional form for the activation functions.

In this neural network model, the objective variable and the explanatory variables are the same as those used in the random forest model. Hence, the number of input nodes is 1,187, and the number of output nodes is 1. In this study, we use Scikit-learn MLPRegressor which implements a multi-layer perceptron (MLP) that trains using backpropagation with identity activation function in the output layer. It uses the square error as the loss function. Therefore, the parameters remained to adjust is the size of hidden layers, including the number of nodes in the hidden layer, and the number of hidden layers, which are also determined by cross-validation.

# 3. CONSTRUCTION OF ATTENUATION RELATIONSHIPS

## 3.1 Construction of the attenuation equation

In this study, 7 iterations are sufficient to determine the regression coefficients. The resulting coefficients from regression are given in table 2, and the resulting equation is

$$\log PGV = -1.541 + 0.648 Mw - 0.00153r - \log(r + 0.0033 * 10^{0.5*Mw}) + 0.00299H + C_i \quad (4)$$

Fig. 2 shows the comparisons of the predicted PGV of the attenuation equation developed in this study with those of Si and Midorikawa (1999), Joyner and Boore (1981), and Molas and Yamazaki (1995) when the magnitude is set to 7; the earthquake source depth is set to 5 km; the station coefficient is set to 0. Fig. 2 shows that it is in harmony with the results of previous studies. Although there are variations among the 4 equations, the attenuation relationship proposed by this study has an intermediate value.

Table 2. Regression coefficients for PGV of the attenuation equation obtained in this study.

| $b_0$ | $b_1$ | $b_2$ | $b_3$ | $b_4$ | $c_1$ | $c_2$ |
|---|---|---|---|---|---|---|
| -1.541 | 0.648 | -0.00153 | -1.00 | 0.00299 | 0.0033 | 0.50 |

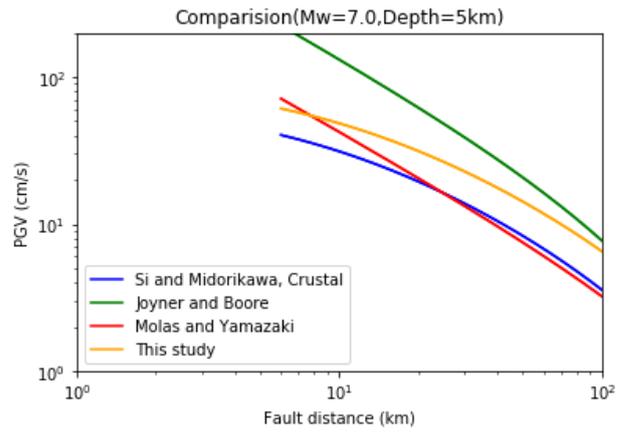

Fig. 2. Comparison of attenuation equations for PGV of this study and previous studies for magnitude 7.0 earthquake with the depth of 5 km.

## 3.2 Construction of machine learning models

In this study, we determine the values of parameters by using GridSearchCV, which is the process of performing hyper parameters tuning in order to determine the optimal values for a given model. Tables 3 and 4 show the parameters of random forest and neural network models obtained by GridSearchCV with CV=5. Fig. 3 shows the predicted attenuation curves for PGV of the attenuation equation, random forest model, and neural network model for the moment magnitudes of 5.0, 6.0 and 7.0 earthquakes with the depth of 5 km. The predictions of machine learning seem to be reasonable and stable like the attenuation equation as demonstrated in Fig. 3. It can also be seen that PGV and the moment magnitude have a clear correlation, although the curve of random forest model is relatively volatile.

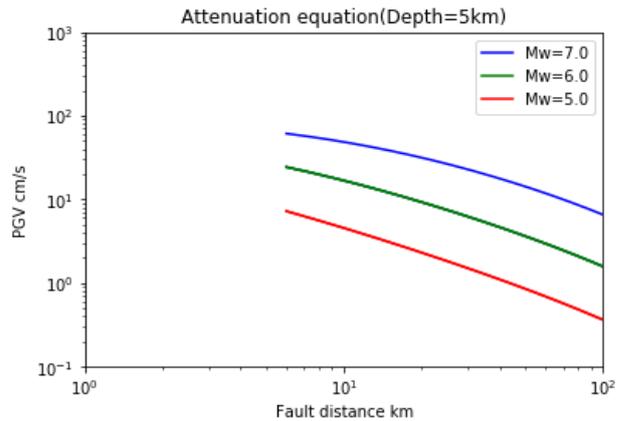

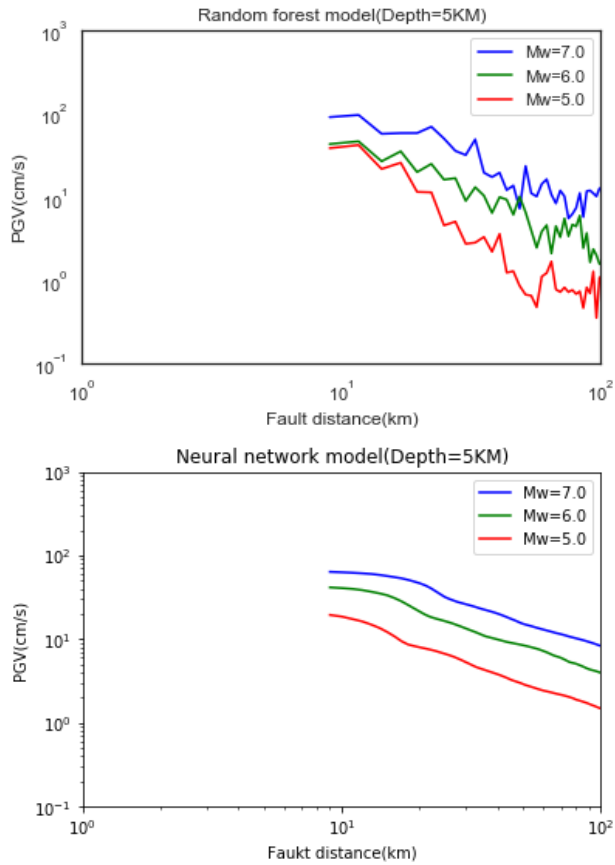

Fig. 3. Predicted attenuation curves for PGV obtained by the regression analysis, random forest, and neural network for the magnitudes of 5.0, 6.0 and 7.0 earthquakes with depth of 5 km.

Table 3. Parameters of random forest model.

| n_estimators | max_depth | max_features |
|---|---|---|
| 1000 | 15 | None |

Table 4. Parameters of neural network model.

| The size of hidden layers |
|---|
| 100-100-100 |

## 4. EVALUATION AND DISCUSSION

After constructing the traditional attenuation equation and the attenuation relationships based on machine learning, we evaluate theirs predictive ability. Ground motion data observed by K-NET and KiK-net during the 4 recent earthquake events are used as the test dataset, which are shown in table 5. Fig. 4 shows the epicenter locations and the moment magnitude. The statistical parameters such as coefficient of determination ($R^2$), mean absolute error (MAE), root mean squared error (RMSE) are used to test the efficacy of the models. The results obtained by the 3 attenuation relationships are shown in Table 6.

Table 5. The list of earthquakes used as the test data.

| No. | Date | Mw | Depth | Number of Records (r<100km) |
|---|---|---|---|---|
| 1 | 2016.04.14 | 6.5 | 11 | 116 |
| 2 | 2016.04.16 | 7.3 | 12 | 128 |
| 3 | 2018.06.18 | 6.1 | 13 | 120 |
| 4 | 2018.09.06 | 6.7 | 35 | 69 |

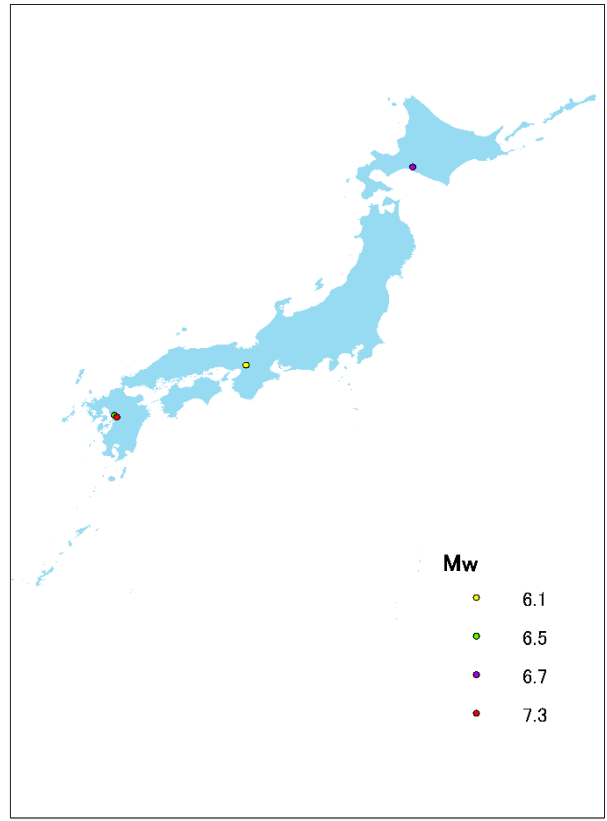

Fig. 4. Locations of epicenter and the moment magnitude of the 4 recent earthquakes.

Table 6. Results of the performance of the 3 models based on the test dataset.

| Predictor | $R^2$ | MAE | RMSE |
|---|---|---|---|
| The April 14 foreshock of 2016 Kumamoto earthquake | | | |
| AE | 0.62 | 4.59 | 7.40 |
| RF | 0.33 | 5.86 | 9.93 |
| NN | 0.48 | 5.01 | 8.69 |
| The April 16 mainshock of 2016 Kumamoto earthquake | | | |
| AE | 0.25 | 12.57 | 18.55 |
| RF | 0.66 | 6.61 | 12.51 |
| NN | -0.82 | 31.68 | 39.28 |
| The 2018 Osaka earthquake | | | |
| AE | 0.71 | 2.37 | 3.67 |
| RF | 0.34 | 4.00 | 5.57 |
| NN | 0.31 | 3.10 | 5.64 |

| | The 2018 Hokkaido Eastern Iburi earthquake. | | | RF | 0.52 | 13.87 | 20.49 |
|---|---|---|---|---|---|---|---|
| | AE | 0.31 | 12.69 | 24.21 | NN | 0.60 | 13.36 | 18.55 |

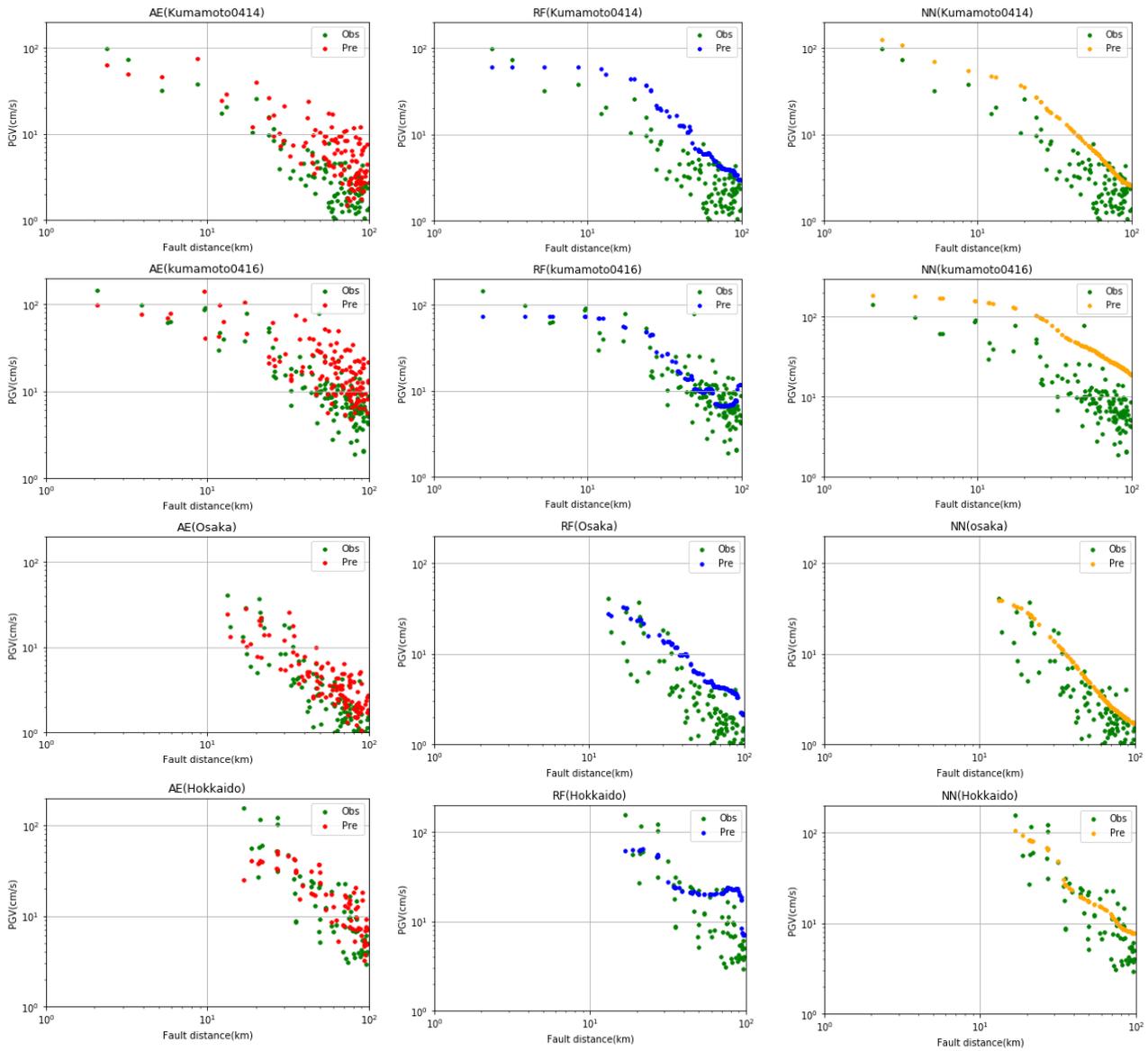

Fig. 5. Observations (Obs) and predictions (Pre) of the 4 recent earthquakes by predictors of the the attenuation equation (AE), random forest (RF), neural network (NN). Red points represent the prediction by the attenuation equation; blue points represent the prediction by the random forest predictor; orange points represent the prediction by the neural network predictor; and green points represent the observed data, respectively.

The observations and predictions for the 4 recent earthquakes in the test dataset are compared to demonstrate the prediction performance of the 3 models in Fig. 5. Figure 5 shows that the overall feature of the observation is reproduced by the machine learning models. In the case of the April 14 foreshock of 2016 Kumamoto earthquake, even if the model of attenuation equation makes a better prediction in the most part, the neural network predictor gets a prediction of over 100 cm/s, whose observation value is also greater than 100 cm/s. It also can be seen that in the case of the April 16 mainshock of 2016 Kumamoto earthquake and the 2018 Hokkaido Eastern Iburi earthquake, the neural network predictor makes the closest predictions on the large observed PGV whose values are greater than 100 cm/s. However, the neural network predictor overestimates the observed PGV, overall. Figure 5 also shows that the random forest model has higher reliability than the attenuation equation at close range. Apart from the case of the April 14 foreshock of 2016 Kumamoto earthquake, the random forest predictor has better performance than the attenuation equation, especially in terms of the prediction of the case of the April 16 mainshock of 2016 Kumamoto earthquake and the 2018 Hokkaido Eastern Iburi earthquake. In the prediction of the maximum value of PGV for the 4 earthquake events, the random forest perfoms better than the attenuation equation except for the case of the April 16 mainshock of 2016 Kumamoto earthquake. However, the random forest predictor trends to overestimate the observed PGV when the shortest distance from the fault is larger than 70 km.

We also check the prediction performance on the training data. Figures 6 shows the relationship between observations and predictions on the training data by the attenuation equation predictor, random forest predictor, neural network predictor, respectively. If there were no trends of underestimation and overestimation, the relationship would be distributed mainly on the diagonal line. As shown in Figs. 6, the neural network predictor is good at predicting large valued PGV, although there is a trend of overestimation when predicting small PGV. The random forest predictor also performs better than the attenuation equation on the prediction of large valued PGV.

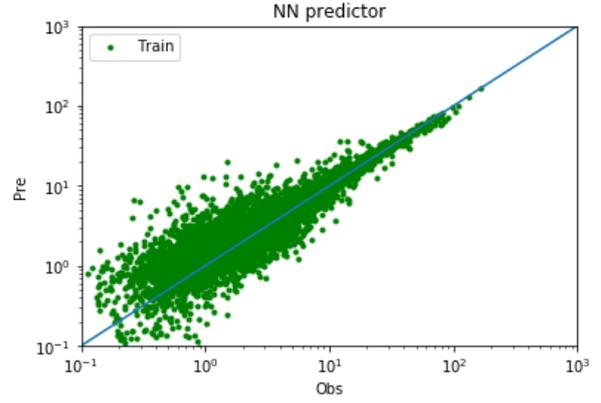

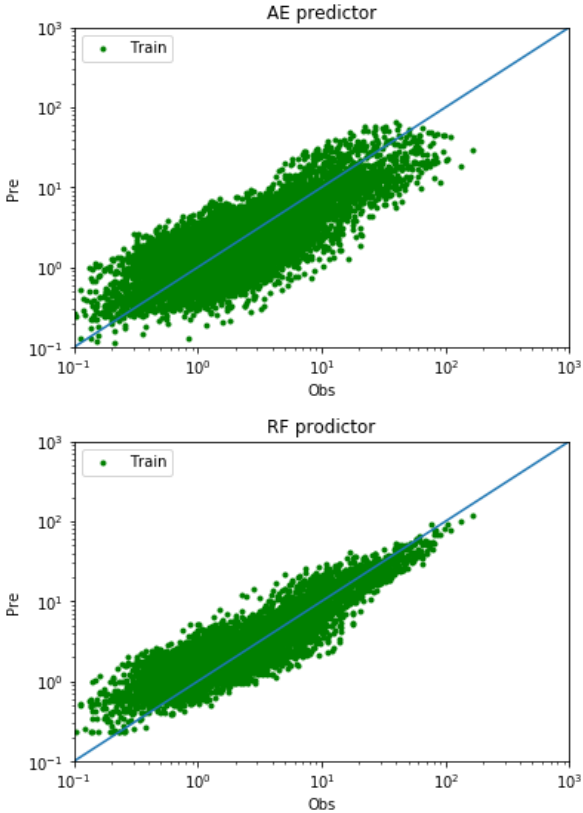

Fig 6. Relationship between observations and predictions in the training dataset by the attenuation equation predictor, the random forest predictor, the neural network predictor, respectively.

In this study, we calculated the station correction factors based on the attenuation relationships of machine learning methods. The station correction factors are effective for evaluating site amplifications. Moreover, in the three-stage regression procedure of the attenuation equation, the station correction factors have been calculated in this study. We make the following procedures in order to calculate the station correction factors based on random forest model and neural network model.

Procedures of random forest model to obtain the station correction factor :
  I. Construct an attenuation relationship for PGV using the moment magnitude, the shortest distance from the fault, and the earthquake source depth as explanatory variables.
  II. Make predictions of PGV based on the model developed by previous step.
  III. The site characteristic value of the observation station can be considered as
  $$\log PGV_{Obs} - \log PGV_{Pre}$$
  IV. Then, the station correction factor ($c_i$) can be calculated by the following equation.
  $$\log PGV_{Obs} - \log PGV_{Pre} = \sum_{i=1}^{N-1} c_i S_{i,j}$$

Where $S_{i,j}$ represents the dummy variable of the observation station which is mentioned before.

Procedures of neural network model to obtain the station correction factor :
  I. Construct an attenuation relationship for PGV using the moment magnitude, the shortest distance from the fault, the earthquake source depth, and the dummy variables of the observation stations as explanatory variables.

II. Fix the moment magnitude, the shortest distance from the fault, the earthquake source depth to a certain value, without changing the dummy variables of the observation stations. Then input the data into the model developed by previous step to get the corresponding PGV which can be regarded as a series of PGV values affected only by the dummy variables of the observation stations.
III. Select the observation station whose station correction factor based on the attenuation equation is 0 as the reference observation station.
IV. Then, the station correction factor for a certain observation station A is

$$c_A = \log PGV_A - \log PGV_{Reference\ station}$$

Figure 7 shows the relationships between the station correction factors calculated in this study and the shear wave velocity averaged over the upper 30 m (AVS30) of the attenuation equation, random forest model, neural network model, respectively. The AVS30 is used for soil classification in the seismic design code in the United States. Figure 7 shows that the station correction factors based on machine learning are correlated with the AVS30, although the coefficients of determination are lower than those of the attenuation equation.

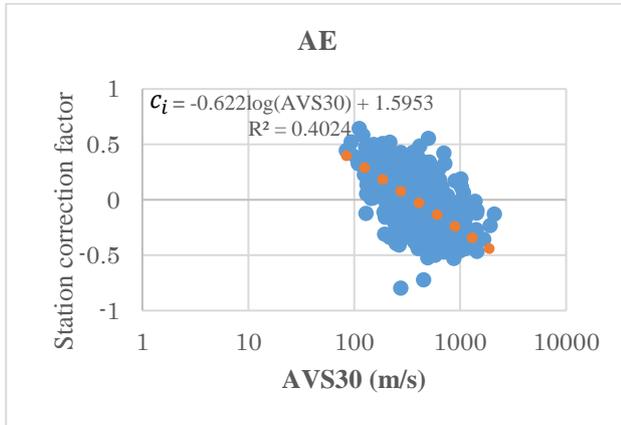

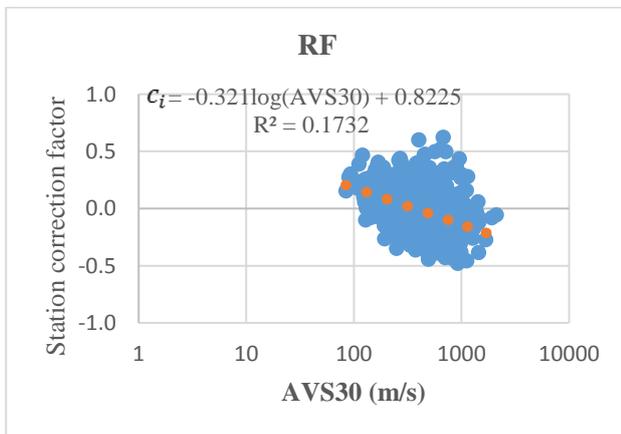

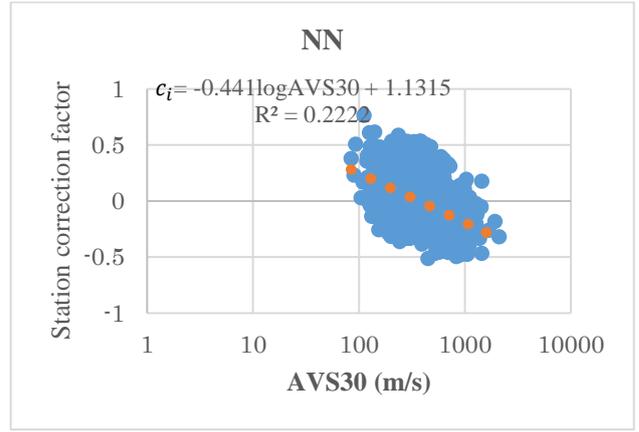

Figure 7. Relationships between the station correction factors and AVS30 based on the attenuation relationship, random forest model, neural network model, respectively.

## 6. CONCLUSION

In this study, a traditional attenuation equation using three-stage regression procedure proposed by Molas and Yamazaki (1995) and the 2 attenuation relationships using random forest and neural network are constructed to predict the PGV. The objective variable and the explanatory variables of the 3 models are set to be the same. The prediction performance of the attenuation equation for the training data is not good as that for the predictors of machine learning. As the machine learning predictor is fully data-driven predictive models when the attenuation equation is formulated. In the case of test data, although the predictor of attenuation equation makes a better prediction for the April 14 foreshock of Kumamoto earthquake. Predictions by the 2 machine learning predictors are improved for the other 3 earthquake events at close range. It seems that machine learning methods improve the reliability of attenuation relationship at close range. In addition, the station correction factors based on machine learning are derived and proved to be correlated with AVS30.

However, there is an overestimation problem for machine learning models. As the shortest distance from the fault increases, both random forest predictor and neural network predictor show overestimation problem. The machine learning models constructed in this study are based on the machine learning library of SK learn. When using GridSearchCV to determine the size of the neural network, it can only be selected from a specific size, which prevents finding the most suitable parameters. In addition, the number of explanatory variables may be one of the reasons caused overestimation. The explanatory variables used in this study contain 1184 dummy variables. According to the random forest model, when there are too many features, the model could identify each training sample, which means overfitting. On the neural network model, irrelevant features affect the fitting of the neural network.